\UseRawInputEncoding

\documentclass[reprint,prd,amsmath,amssymb,aps,floatfix,twocolumn,nofootinbib]{revtex4-1}

\usepackage[english]{babel}
\usepackage{graphicx}
\usepackage{amssymb,amsmath}
\usepackage{xcolor}
\usepackage{hyperref}
\usepackage{aas_macros}

\begin{document}


\title{The first minutes of a binary-driven hypernova}

\author{L.~M.~Becerra,$^{9,1}$ R.~Moradi,$^{1,3,6}$ J.~A.~Rueda,$^{1,2,3,4,5}$ R. Ruffini,$^{1,3,7,8}$ Y. Wang$^{1,3,6}$}

\affiliation{$^{1}$ICRANet, Piazza della Repubblica 10, I-65122 Pescara, Italy}
\affiliation{$^{2}$ICRANet-Ferrara, Dipartimento di Fisica e Scienze della Terra, Universit\`a degli Studi di Ferrara, Via Saragat 1, I-44122 Ferrara, Italy}
\affiliation{$^{3}$ICRA, Dipartimento di Fisica, Sapienza Universit\`a  di Roma, Piazzale Aldo Moro 5, I-00185 Roma, Italy}
\affiliation{$^{4}$Dipartimento di Fisica e Scienze della Terra, Universit\`a degli Studi di Ferrara, Via Saragat 1, I-44122 Ferrara, Italy}
\affiliation{$^{5}$INAF, Istituto di Astrofisica e Planetologia Spaziali, Via Fosso del Cavaliere 100, I-00133 Rome, Italy}
\affiliation{$^{6}$INAF, Osservatorio Astronomico d'Abruzzo,Via M. Maggini snc, I-64100, Teramo, Italy}
\affiliation{$^{7}$Universit\'e de Nice Sophia-Antipolis, Grand Ch\^ateau Parc Valrose, Nice, CEDEX 2, France}
\affiliation{$^{8}$INAF, Viale del Parco Mellini 84, I-00136 Rome, Italy}
\affiliation{$^{9}$Escuela de F\'isica, Universidad Industrial de Santander, A.A.678, Bucaramanga, 680002, Colombia }

\email{laura.becerra7@correo.uis.edu.co, rahim.moradu@icranet.org\\ jorge.rueda@icra.it, ruffini@icra.it, yu.wang@icranet.org}

\date{\today / Received date /
Accepted date}

\begin{abstract}
We simulate the first minutes of the evolution of a binary-driven hypernova (BdHN) event, with a special focus on the associated accretion processes of supernova (SN) ejecta onto the newborn neutron star ($\nu$NS) and the NS companion. We calculate the rotational evolution of the $\nu$NS and the NS under the torques exerted by the accreted matter and the magnetic field. We take into account general relativistic effects {through effective models for the NSs binding energy and the specific angular momentum transferred by the accreted matter.} We use realistic hypercritical accretion rates obtained from three-dimensional smoothed-particle-hydrodynamics (SPH) numerical simulations of the BdHN for a variety of orbital periods. We show that the rotation power of the $\nu$NS has a unique double-peak structure while that of the NS has a single peak. These peaks are of comparable intensity and can occur very close in time or even simultaneously depending on the orbital period and the initial angular momentum of the stars. We outline the consequences of the above features in the early emission and their consequent observation in long gamma-ray bursts.
\end{abstract}


\maketitle


\section{Introduction}\label{sec:1}

{
The phenomenological classification of gamma-ray bursts (GRBs) is based on the observed time it takes to release $90\%$ of the total isotropic energy ($E_{\rm iso}$) in the gamma-rays prompt emission, $T_{90}$. Long GRBs are those with $T_{90}>2$~s and short GRBs are the sources with $T_{90}<2$~s \cite{1981Ap&SS..80....3M, 1992grbo.book..161K, 1992AIPC..265..304D, 1993ApJ...413L.101K, 1998ApJ...497L..21T}. The Burst And Transient Source Experiment (BATSE) on board the COMPTON Gamma-Ray Observatory (CGRO) showed the isotropic distribution of GRBs in the sky, which suggests their extragalactic origin \cite{1992Natur.355..143M}. The BeppoSAX satellite launched allowed the follow-up of the GRB emission leading to the discovery of a long-lasting X-ray afterglow with the first case of GRB 970228 \cite{1997Natur.387..783C}. BeppoSAX improved the GRB localization to arcminutes resolution, which allowed the detection of the optical counterparts and host galaxies by earth-based telescopes. These observations led to measuring the cosmological redshift of GRBs. More afterglows were detected, and the cosmological distances of $\approx5$--$10$~Gpc confirmed the GRB cosmological origin. These measurements confirmed the (long-time suspected) great energy release of GRBs, $E_{\rm iso}\approx 10^{50}$--$10^{54}$~erg.
}

{
It was soon reached the consensus that the huge energetics involved in both short and long GRBs imply they are related to the process of gravitational collapse at the end of massive stars, i.e., processes involving neutron stars (NSs) and/or black holes (BHs). For short bursts, NS-NS and/or NS-BH binary mergers were proposed (see e.g. the pioneering works \cite{1986ApJ...308L..47G, 1986ApJ...308L..43P, 1989Natur.340..126E, 1991ApJ...379L..17N}). For long bursts, the picture of a \textit{collapsar} \cite{1993ApJ...405..273W}, the core-collapse of a single massive star leading to a BH (or a magnetar) surrounded by a massive accretion disk, has become the traditional GRB model (see, e.g., \cite{2002ARA26A..40..137M, 2004RvMP...76.1143P}, for reviews). The traditional model for the prompt emission of both short and long GRBs follows the dynamics of a \textit{fireball}, an optically thick plasma of electron-positron ($e^-e^+$) pairs and photons in equilibrium with a baryonic plasma \cite{1978MNRAS.183..359C, 1986ApJ...308L..43P, 1986ApJ...308L..47G, 1991ApJ...379L..17N, 1992ApJ...395L..83N}. The current version of the traditional model of GRB assumes the fireball expands in a collimated relativistic jet expanding with Lorentz factors $\Gamma \sim 10^2$--$10^3$ \cite{1990ApJ...365L..55S, 1992MNRAS.258P..41R, 1993MNRAS.263..861P, 1993ApJ...415..181M, 1994ApJ...424L.131M}. The internal shock produces the prompt emission, and the external shock generates the afterglow by interacting with the interstellar medium producing synchrotron radiation, and very-high-energy (VHE) emission by synchrotron self-Compton \cite{2002ARA26A..40..137M, 2004RvMP...76.1143P, 2019Natur.575..455M, 2019Natur.575..448Z}. There have been additional details, modifications and/or extensions made to the above main picture, and we refer the reader to the recent comprehensive book by \citet{2018pgrb.book.....Z} for more details on the latest developments of the traditional GRB model.
}

{
The optical follow-up of the afterglow guided by the GRB localization by BeppoSAX (then extended by the Neil Gehrels Swift Observatory \textcolor{green}{\citep{2005SSRv..120..143B,2005SSRv..120..165B,2005SSRv..120...95R}}) led to another great discovery: the association of long GRBs with type Ic supernovae (SNe). The first evidence of such an association was the temporal and spatial coincidence of GRB 980425 and SN 1998bw \cite{1998Natur.395..670G}. Many additional associations followed, confirming the GRB-SN connection \cite{2006ARA&A..44..507W, 2011IJMPD..20.1745D, 2012grb..book..169H}. There have been attempts to overcome the natural drawback of the extreme requirement of the gravitational collapse of a massive star to form a collapsar, the jetted fireball, and an SN explosion. Some models propose that an efficient neutrino emission from the accretion disk might power a successful SN explosion \cite{1999ApJ...524..262M} or a beamed outflow/wind that hosts the nucleosynthesis of the nickel required to explain the optical SN (see, e.g., \cite{2005ApJ...629..341K, 2012ApJ...744..103M, 2012ApJ...750..163L}).
}

{
Having recalled the generalities of the traditional model of GRBs, we turn now to the alternative scenario based on a binary progenitor. First, we recall the seminal work of \citet{1999ApJ...526..152F} that, from a binary stellar evolution viewpoint, pointed out a variety of binaries that can lead to GRB events. Second, long GRBs and SNe are characterized by very different energetics, the latter in the range $10^{49}$--$10^{51}$~erg, and the former in the range $10^{49}$--$10^{54}$~erg. The high GRB energetics point to the gravitational collapse to a stellar-mass BH, while a SN originate in the core-collapse of a massive star forming a NS. The formation of a BH in core-collapse SN is discarded by the low observed masses of pre-SN progenitors, $\lesssim 18~M_\odot$ \cite{2015PASA...32...16S}, which are unable to lead to direct collapse to a BH (see, \cite{2009ARA&A..47...63S, 2015PASA...32...16S} {, although the threshold mass for BH formation is not sharply defined and may depend on several physical pre-SN star properties \cite{2012ApJ...757...69U}).} From this point of view, it seems unlikely that the GRB and the SN can both originate from the single-star progenitor. One of the most compelling reasons for the quest for a binary GRB progenitor arises from their association with SNe of type Ic, i.e., that lack hydrogen (H) and helium (He) in their spectra. From theory and observations, the most accepted view is that SNe Ic are produced by bare He, carbon-oxygen (CO), or Wolf-Rayet (WR) stars whose hydrogen and helium envelopes have been stripped during their evolution (see, e.g., \cite{2011MNRAS.415..773S, 2020MNRAS.492.4369T}). The stripped-envelope He/CO/WR are thought to form tight binaries with a compact-star companion (e.g. a NS) that helps them to get rid of its H/He layers through multiple mass-transfer and common-envelope phases (see, e.g., \cite{1988PhR...163...13N, 1994ApJ...437L.115I, 2007PASP..119.1211F, 2010ApJ...725..940Y, 2011MNRAS.415..773S, 2015PASA...32...15Y, 2015ApJ...809..131K}, and Section \ref{sec:II} for further details). 
}

\begin{figure*}
    \centering
    \includegraphics[width=0.9\hsize,clip]{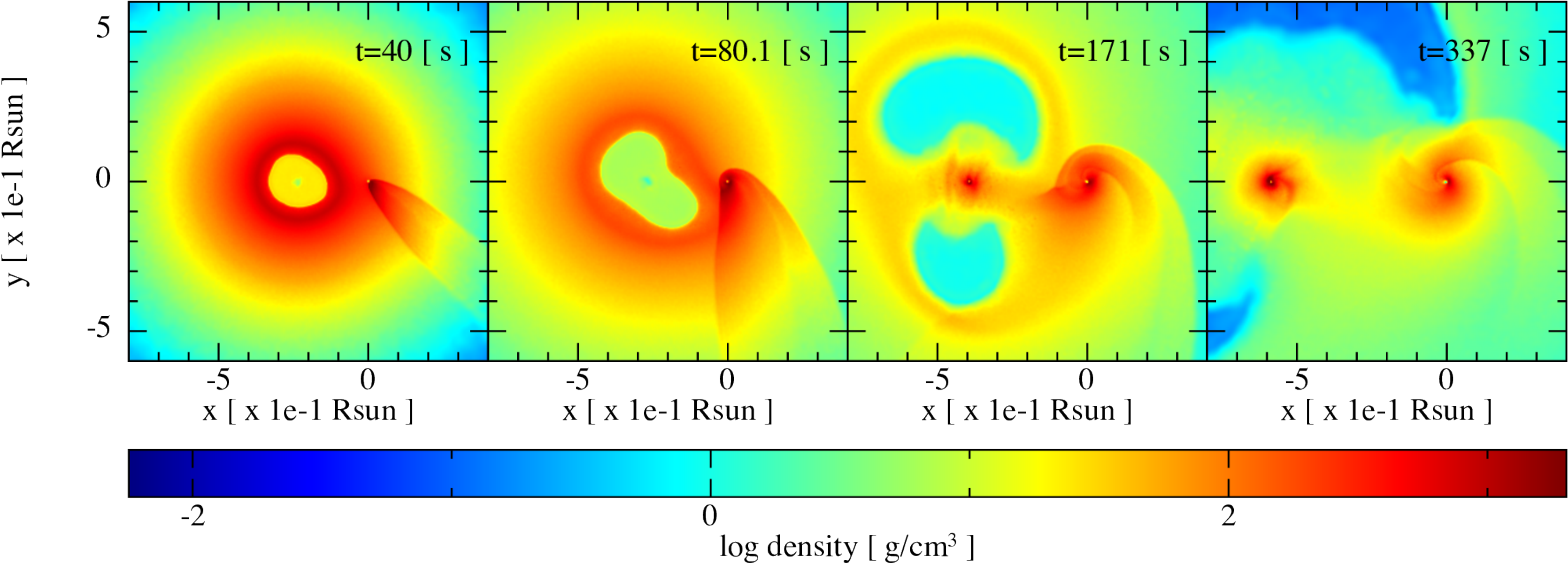}\\
    \includegraphics[width=0.9\hsize,clip]{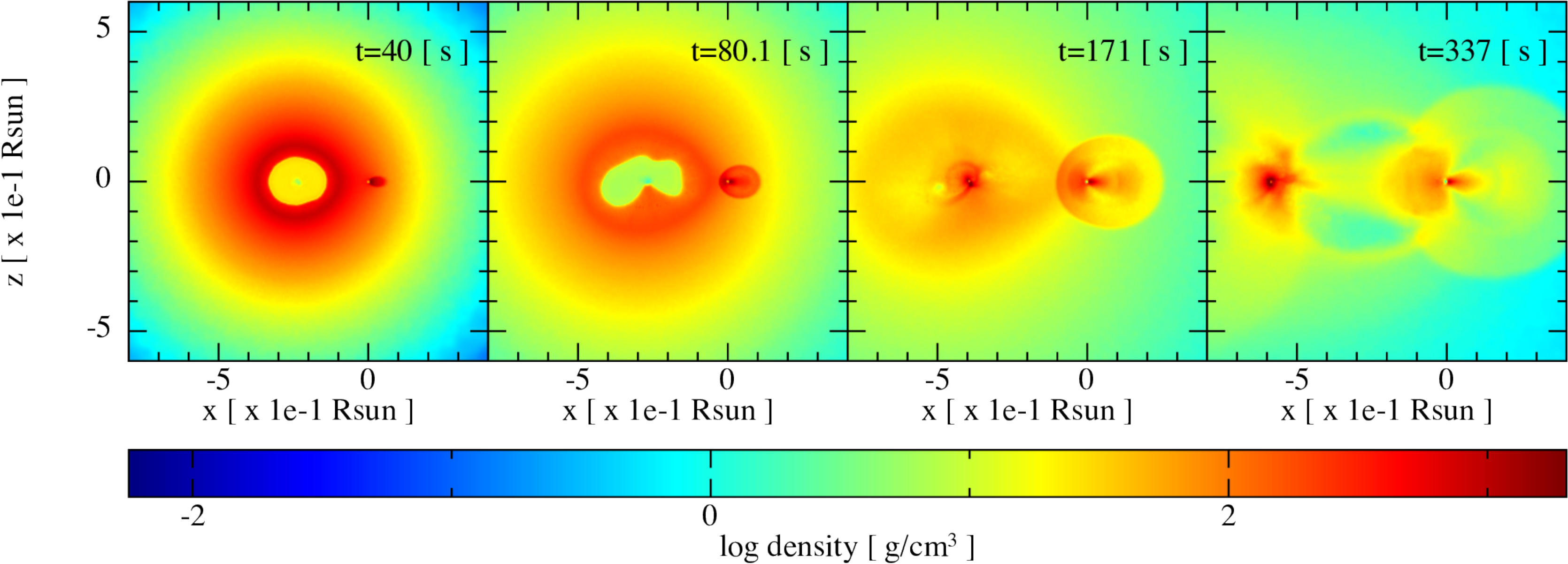}
    \caption{SPH simulation of a BdHN I: model ``30m1p1eb'' of Table 2 in \cite{2019ApJ...871...14B}. The binary progenitor is composed of a CO star of $\approx 9 M_\odot$, produced by a ZAMS star of $30 M_\odot$, and a $2 M_\odot$ NS companion. The orbital period is $\approx 6$ minutes. Each frame, from left to right, corresponds to selected increasing times with $t = 0$ s the instant of the SN shock breakout. The upper panel shows the mass density on the equatorial plane and the lower panel the plane orthogonal to the equatorial one. The reference system is rotated and translated to align the x-axis with the line joining the binary components. The origin of the reference system is located at the NS companion position. The first frame corresponds to $t = 40$ s, and it shows that the particles entering the NS capture region form a tail behind the star. These particles then circularize around the NS, forming a thick disk that is already visible in the second frame at $t = 80$ s. Part of the SN ejecta is also accreted by the $\nu$NS as is appreciable in the third frame at $t=171$ s. At $t = 337$ s (about one orbital period), a disk structure has been formed around the $\nu$NS and the NS companion.
    This figure has been produced with the SNsplash visualization program \cite{2011ascl.soft03004P}.
    }
    \label{fig:3DSPH}
\end{figure*}

{
The binary-driven hypernova (BdHN) model of long GRBs follows the natural fate of some stripped-envelope binaries. The GRB progenitor is a CO-NS binary at the end of the thermonuclear evolution of the CO star, i.e., at the second core-collapse SN event in the evolution of the binary (the first SN formed the NS companion; see Sec. \ref{sec:II} for details on the binary evolutionary path). The BdHN is rooted in the concept of induced gravitational collapse (IGC) that can occur when the CO star undergoes SN in presence of the NS companion \cite{2012ApJ...758L...7R}, and the different emissions observed in the GRB are explained through a sequence of physical processes following the SN explosion (see, e.g., \cite{2012ApJ...758L...7R, 2012A&A...548L...5I, 2014ApJ...793L..36F, 2015PhRvL.115w1102F, 2015ApJ...812..100B, 2016ApJ...833..107B, 2019ApJ...871...14B}).} The sequence of physical processes is as follows (see Fig. \ref{fig:3DSPH}). The gravitational collapse of the iron core of the CO star leads to the formation of a newborn NS ($\nu$NS) at its center and ejects the outer layers of the star. The ejecta triggers an accretion process onto the NS companion, while matter fallback also occurs leading to an accretion process onto the $\nu$NS. Both accretion processes proceed at hypercritical (i.e., highly super-Eddington) rates thanks to a copious neutrino emission \cite{2016ApJ...833..107B, 2018ApJ...852..120B}. In compact binaries with orbital periods of a few minutes, the hypercritical accretion onto the NS companion brings it the critical mass inducing its gravitational collapse and forming a rotating (Kerr) BH. We have called these systems BdHN of type I (hereafter BdHN I). In less compact binaries, the NS companion does not reach the critical mass and holds stable as a more massive and fast rotating NS. We have called these systems BdHN of type II (hereafter BdHN II). {Therefore, if the binary is not disrupted by the explosion, the BdHN scenario contemplates two possible fates, the formation of a NS-NS (in BdHN II), or the formation of a NS-BH (in BdHN I). Only the former fate has been considered in stripped-envelope binaries (see Sec. \ref{sec:II}).}

{
We recall in Sec. \ref{sec:II} the sequence of physical processes that occur in BdHN I and II triggered by the SN explosion in the CO-NS progenitor binary. Those processes lead to specific observables (episodes) that can be identified in the data (multiwavelength light-curves and spectra) of high-energetic (BdHN I) and low-energetic (BdHN II) GRBs. In-depth time-resolved analyses have led to the interpretation of the GRB data in terms of the above physical episodes in both BdHN I and II (see, e.g., \cite{2021PhRvD.104f3043M, 2021MNRAS.504.5301R, 2020ApJ...893..148R, 2019ApJ...874...39W, 2019ApJ...886...82R}, and references therein). 
}

In this article, we focus on the possible observable emission by the $\nu$NS and the NS companion during their individual hypercritical accretion processes in the first minutes of evolution following the core-collapse of the CO star leading to the SN explosion, and address its possible observables in long GRBs. {The relevance of this task is boosted by the recent results obtained in the detailed time-resolved analysis of GRB 190829A \cite{2022ApJ...936..190W}. This source is a low-luminosity GRB, hence interpreted as a BdHNe II. The prompt emission of GRB 190829A shows a double-peak structure as expected from the emissions due to accretion process onto the $\nu$NS and the NS companion \cite{2019ApJ...871...14B}. The additional properties of the BdHN model, e.g., the multiwavelength afterglow by synchrotron radiation of the SN ejecta powered by the $\nu$NS, also fit the observational data of this source.
}

{The above results on the successful qualitative and quantitative interpretation of GRB 190829A encourage us to enlarge our knowledge on the early accretion process in BdHNe by exploring as much as possible the system parameters. In this line, we recall that \citet{2019ApJ...871...14B} presented a comprehensive analysis of the accretion process onto both NSs exploring a wide window of values for some system parameters such as the SN explosion energy, orbital period, initial NS companion mass, CO star mass, asymmetric SN explosions, and three different NS equation of state (EOS). These simulations assume the NSs are initially non-rotating and aim to determine the fate of the NSs. Namely, they evaluate if the NSs reach the critical mass for gravitational collapse and form a BH. We aim here to assess the previously uncounted effect of an initially rotating NS companion. In particular, we evaluate how an initially non-zero angular momentum of the NS companion affects the rate at which the $\nu$NS and the NS gain gravitational mass, angular momentum, and rotational power. The latter is a proxy for the power releasable by the stars in the early BdHN evolution before the BH formation. Based on these results, we discuss the possible observational features of the above process in the data of low-energetic and high-energetic GRBs.}

{The article is organized as follows. In Sec. \ref{sec:II}, we summarize the physical processes leading to the GRB observables and the stellar evolution formation channel of BdHNe. Section  \ref{sec:III} presents the numerical simulations of the SN explosion in the CO-NS binary used in this article, with emphasis on the estimation of the accretion rates onto the $\nu$NS and the NS companion. We present in Sec. \ref{sec:IV} the treatment to calculate the evolution of the NSs structure parameters with focus on the rotational evolution under the action of accretion and magnetic torques. In Sec. \ref{sec:V}, we calculate the rotational energy gained by the stars during the hypercritical accretion for a variety of initial conditions, and discuss the relevant features for GRB observations. Finally, in Sec. \ref{sec:VI} we draw the conclusions of this article.}

\section{Physical processes, observables, and evolutionary path of BdHNe}\label{sec:II}

Before entering into details of the calculation, we recall the emission episodes of a BdHN event and how they are associated with the observed emissions in a long GRB.

\subsection{Physical processes and related observables}

First, as we shall show in this article, the SN explosion and the hypercritical accretion onto the $\nu$NS and the NS companion can be observed as precursors to the prompt gamma-ray emission (see, e.g., \cite{2016ApJ...833..107B, 2019ApJ...874...39W}).

In the case of BH formation (BdHN I), the gravitomagnetic interaction of the newborn Kerr BH with the surrounding magnetic field inherited from the collapsed NS induces an electric field. This system is what we have called the \textit{inner engine} of the high-energy emission of long GRBs \cite{2019ApJ...886...82R, 2020EPJC...80..300R, 2021A&A...649A..75M, 2021MNRAS.504.5301R}.

The induced electric field is initially overcritical leading to an electron-positron ($e^+e^-$) pair plasma. The plasma self-accelerates to ultrarelativistic velocities and once it becomes transparent its gamma-rays emission is observed as the GRB ultrarelativistic prompt emission (UPE) phase (see \cite{2021PhRvD.104f3043M, 2022EPJC...82..778R}, for details). The $e^+e^-$ plasma loaded with baryons from the SN ejecta, expand through the ejecta and as it gets transparent lead to the gamma- and X-ray flares observed in the early afterglow \cite{2018ApJ...852...53R}. 

The electric field accelerates to ultrarelativistic velocities the electrons from the matter surrounding the BH. Along the BH rotation axis the electric and magnetic field are parallel, so there are no significant radiation losses, implying that electrons gain a kinetic energy equal to the electric potential energy difference from the acceleration point to infinity. Electrons can reach energies of up to $10^{18}$~eV and protons of up to $10^{21}$~eV \cite{2020EPJC...80..300R}, hence contributing to ultra high-energy cosmic rays (UHECRs). Off-polar axis, synchrotron radiation losses occur leading to the observed GeV emission of long GRBs \cite{2019ApJ...886...82R, 2020EPJC...80..300R, 2021A&A...649A..75M, 2022ApJ...929...56R}.

Synchrotron radiation by relativistic electrons in the ejecta expanding in the magnetized medium provided by the $\nu$NS magnetic field, and powered by the $\nu$NS rotational energy, explains the afterglow emission in the X-rays, optical, and radio wavelengths \cite{2018ApJ...869..101R, 2019ApJ...874...39W, 2020ApJ...893..148R}. Because this synchrotron afterglow depends only on the $\nu$NS, it is present both in BdHN I and II.

Finally, there is the emission observed in the optical band powered by the energy release of nickel decay (into cobalt) in the SN ejecta. We refer to \cite{2021IJMPD..3030007R, 2021ARep...65.1026R, 2019Univ....5..110R} for recent reviews on the BdHN scenario of long GRBs and the related physical phenomena. 

\subsection{Evolutionary path}

{
Possible binary evolution paths for the formation of the CO-NS binaries of BdHNe have been discussed in \cite{2015PhRvL.115w1102F}. The natural evolutionary paths are those conceived for the formation of ultra-stripped binaries, mainly introduced in the literature for the explanation of the population of binary neutron stars and low-luminosity and/or rapid-decay-rate SNe \cite{2013ApJ...778L..23T, 2015MNRAS.451.2123T}. The evolution starts from two massive stars. The first core-collapse SN of the primary star forms a NS. After that, the system undergoes a series of mass transfer phases, ejecting both the hydrogen and helium shells of the secondary to produce a binary composed of a massive CO star and a NS.
}

{
The traditional picture assumes that the second SN explosion in the binary evolution, when the iron core of the CO star undergoes gravitational collapse, forms an NS-NS binary (see, e.g., \cite{2015MNRAS.451.2123T, 2017ApJ...846..170T, 2018Sci...362..201D}). The BdHN model explores the possibility that for short orbital period binaries, the NS companion of the exploding CO star can accrete enough mass to reach the critical mass and form a Kerr BH. Numerical simulations include most of the relevant physical processes occurring in the cataclysmic event (see Sec. \ref{sec:III} for details). In those cases, the explosion generates an NS-BH if the binary keeps bound \cite{2015PhRvL.115w1102F}.} Our numerical simulations of the BdHN scenario show three possible fates for the binaries: NS-NS, NS-BH, or binary disruption \cite{2019ApJ...871...14B}.

{
Simulations of the stellar evolution of massive binaries with detailed physics are still under development. Most stellar evolution simulations have focused on single stars, and most of the existing simulations of binary evolution do not self-consistently account for possible effects of the binary interactions on the thermonuclear and mass loss of the binary components (see discussion in \cite{2021ApJ...920L..36J, 2021ApJ...916L...5V}). These latter works represent a step toward self-consistent stellar evolution models leading to ultra-stripped binaries, and possibly allowing for more compact binaries where the BdHN process occurs. Current simulations have lead to CO/He-NS binaries with orbital periods as short as $50$ minutes (e.g. \cite{2015MNRAS.451.2123T}), which is close to the orbital periods we consider here (see e.g. Fig. \ref{fig:Mdots_Porb}).
}

{
Because of the rareness of the GRB phenomenon, consistent with the short orbital periods required for BH formation in BdHNe, we expect these CO-NS binaries to be a small subset of the ultra-stripped binaries. Since $0.1\%$-–$1\%$ of the total SN Ibc are expected to be ultra-stripped binaries \cite{2015MNRAS.451.2123T}, we estimate that only $\approx 0.01\%$-–$0.1\%$ of ultra-stripped binaries are needed to explain the observed population of BdHNe \cite{2015PhRvL.115w1102F}.
}

\section{Simulation of the BdHN early evolution}\label{sec:III}

\begin{figure*}[htb]
    \centering
   \includegraphics[width=0.49\hsize,clip]{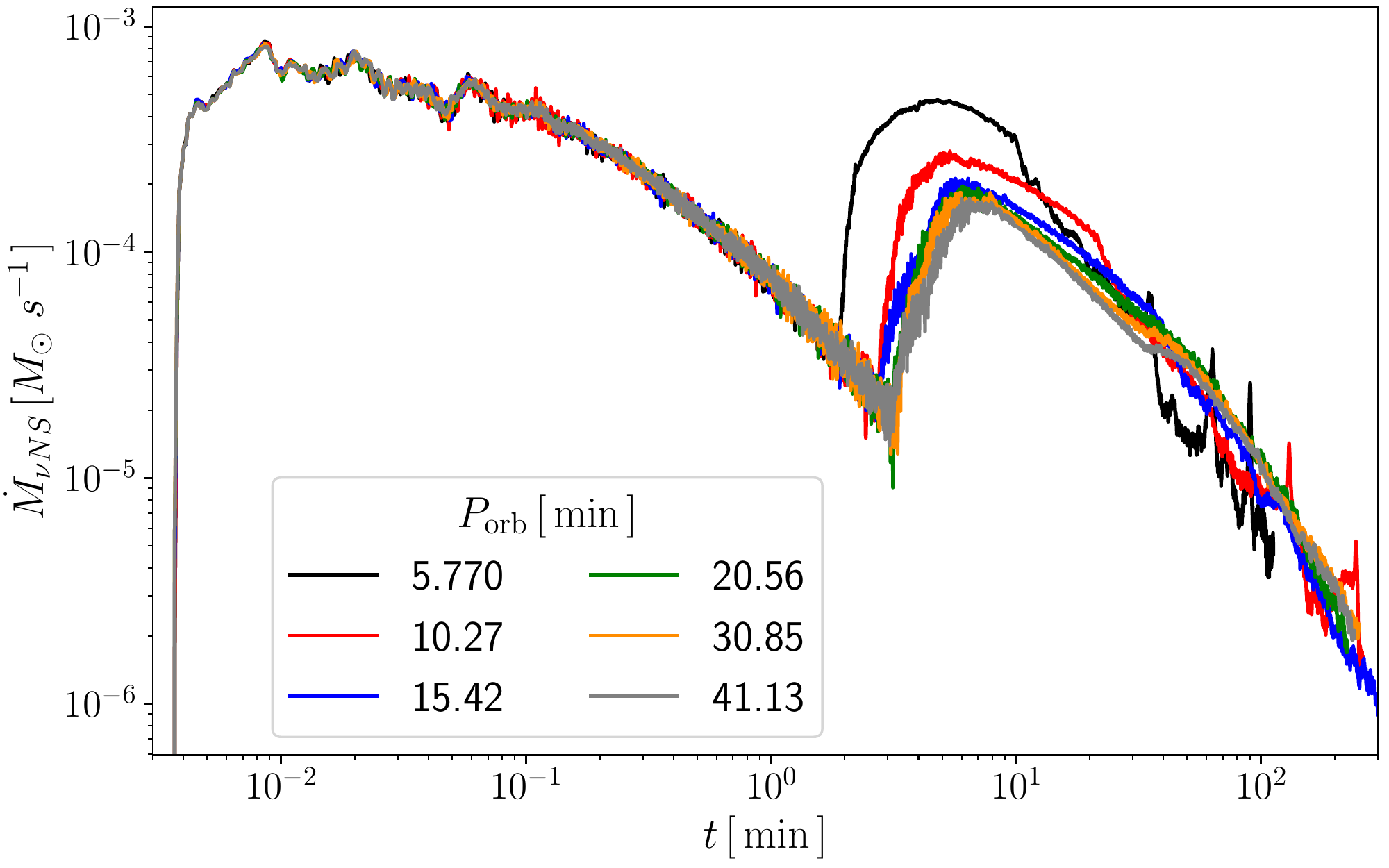}
   \includegraphics[width=0.49\hsize,clip]{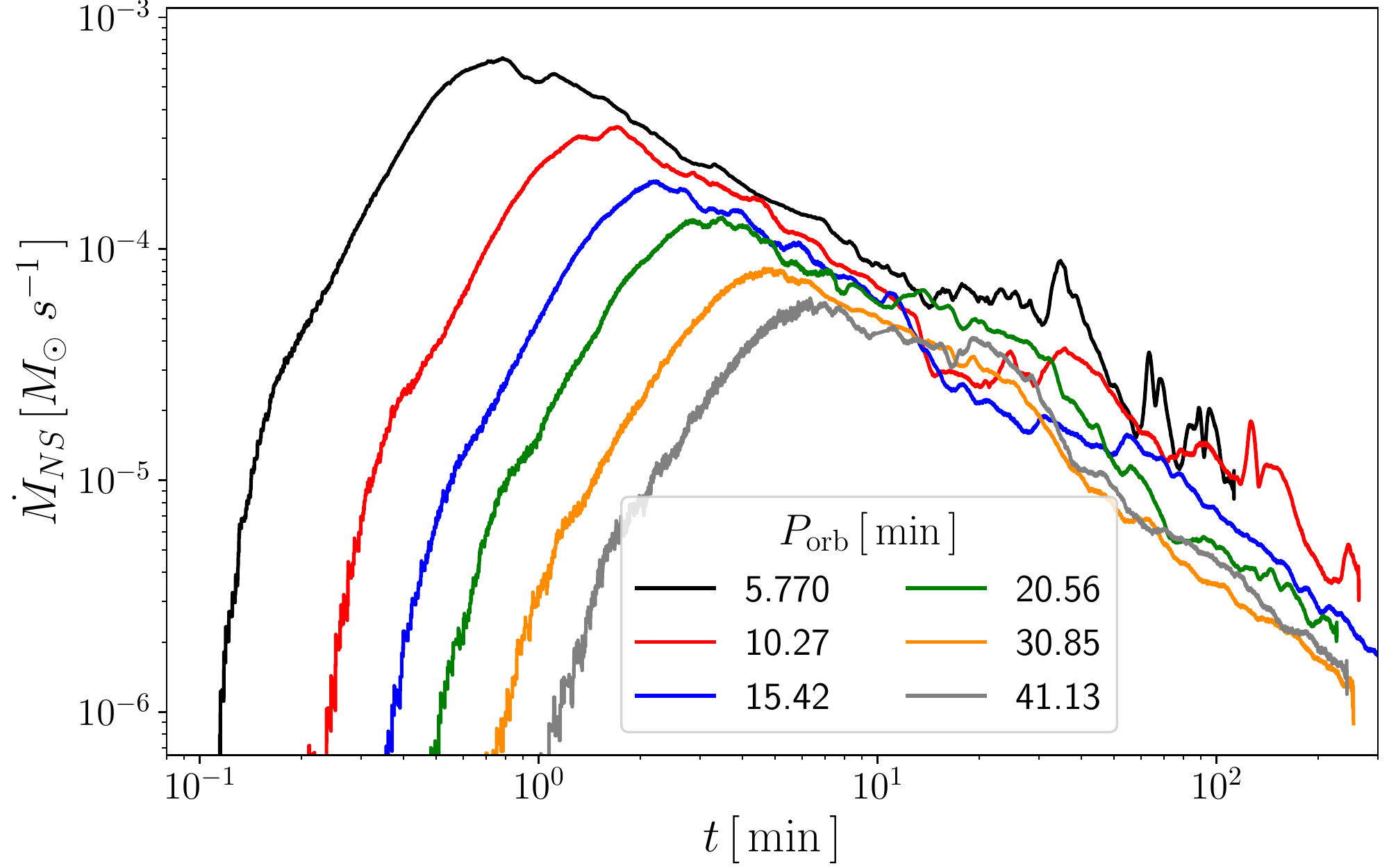}
    \caption{Accretion rate onto the $\nu$NS (left) and onto the NS companion (right) as a function of time, obtained from SPH simulations of BdHNe with different orbital periods.}
    \label{fig:Mdots_Porb}
\end{figure*}

We first obtain from numerical simulations a realistic time evolution of the accretion rate onto the $\nu$NS and the NS companion. We perform smoothed-particle-hydrodynamics (SPH) simulations with the \textit{SNSPH} code adapted to the binary system progenitor of the BdHN presented in \cite{2019ApJ...871...14B}. {This Newtonian, three-dimensional (3D) Langrangian code calculates the evolution of the position, momentum (linear and angular), and thermodynamics (pressure, density, temperature) of the particles. The code calculates the accretion rate by counting the particles that become gravitationally captured by either NS. The Newtonian scheme suffices for this task since the size of the gravitational capture region (i.e., the Bondi-Hoyle radius) for the system parameters is hundreds to thousands bigger than the Schwarzschild radius of the NSs \cite{2019ApJ...871...14B}. Having obtained the rate of particles (i.e., baryonic mass) that are gravitationally captured by the NSs, i.e., $\dot{M}_{\nu \rm NS}^{\rm cap}$ and $\dot{M}_{\rm NS}^{\rm cap}$, we calculate the NSs gravitational mass and angular momentum evolution including general relativistic effects as described in Sec. \ref{sec:III}.}

Figure \ref{fig:3DSPH} shows snapshots of the mass density of the SN ejecta in the y-x plane, the binary equatorial plane, and the z-x plane at different times. In this simulation, the mass of the CO star, just before its collapse, is around $8.89~M_\odot$. This pre-SN configuration is obtained from the thermonuclear evolution of a zero-age-main-sequence (ZAMS) star of $M_{\rm ZAMS}=30~M_\odot$. The NS companion has a mass of $2~M_\odot$, and the orbital period before the SN explosion is of $5.77$~min. The latter is the shortest orbital period that the system can have without triggering Roche-lobe overflow (see, e.g., \cite{2014ApJ...793L..36F}).

The SPH simulation starts when the SN shock front reaches the CO star surface, i.e., we mapped to a 3D-SPH configuration, the 1D core-collapse supernova simulation of \cite{2018ApJ...856...63F}. At this moment, the collapse of the CO star has formed a $\nu$NS of $1.75~M_\odot$, and around $7.14~M_\odot$ is ejected by the SN explosion. In the simulation, the $\nu$NS and NS companion are modeled as point-masses, and interact only gravitationally with the SN-particles and between them. We allow these two point particles to increase their mass by accreting other particles from the SN material following the algorithm described in \cite{2019ApJ...871...14B}.

Figure \ref{fig:3DSPH} shows that the SN ejecta which is gravitationally captured by the NS companion forms first a tail behind the star, and then circularize around it forming a thick disk. At the same time, the particles from the innermost layers of the SN ejecta that were not able to escape from the $\nu$NS gravitational field, fallback and are accreted by the $\nu$NS. After a few minutes, part of the material in the disk around the NS companion is also attracted by the $\nu$NS, producing an enhancement of the accretion process onto the $\nu$NS. The hydrodynamics of the matter infalling and accreting onto a NS at hypercritical rates has been extensively studied in different astrophysics contexts taking into account details on the neutrino emission, e.g., fallback accretion in SN \cite{1972SvA....16..209Z,1996ApJ...460..801F,2006ApJ...646L.131F,2009ApJ...699..409F}, accreting NS in X-ray binaries \cite{1973PhRvL..31.1362R}, and for the case of BdHNe, we refer to \cite{2014ApJ...793L..36F,2016ApJ...833..107B,2018ApJ...852..120B} for details. {The latter includes a formulation in a general relativistic background, and account for neutrino flavour oscillations. The relevant, not obvious result is that these simulations show that the NS can indeed accrete the matter at the hypercritical rate at which baryonic mass from the SN ejecta falls into the gravitational capture region of the NS. Therefore, we assume in this article that the accretion rate inferred with the SPH code as described in Section \ref{sec:II} as the effective baryonic mass accretion rate onto the NS, i.e., we assume $\dot{M}_{b, \nu \rm NS}  = \dot{M}_{\nu \rm NS}^{\rm cap}$ and $\dot{M}_{b, \rm NS} = \dot{M}_{\rm NS}^{\rm cap}$.}

Figure \ref{fig:Mdots_Porb} shows the accretion rate onto the $\nu$NS and the NS companion obtained from SPH simulations for selected orbital periods that cover a range of BdHN I and II. The accretion rate onto the $\nu$NS shows two prominent peaks. The second peak of the fallback accretion rate onto the $\nu$NS is a unique feature of BdHNe because, as explained above (see, also, \cite{2019ApJ...871...14B}, for additional details), it is caused by the influence of NS companion. The accretion rate onto the NS companion shows a single-peak structure, accompanied by additional peaks of smaller intensity and shorter timescales, more evident in binaries with short orbital periods. Such small peaks are produced by episodes of higher and lower accretion that occur as the NS companion orbits across the ejecta and find regions of higher and lower density.

\section{Neutron star rotational evolution}\label{sec:IV}

We calculate the evolution of the $\nu$NS and the NS companion gravitational mass and angular momentum following the formalism described in \cite{2019ApJ...871...14B}. {At every time, we describe the NS as a rigidly rotating configuration described by a stationary, axisymmetric metric fulfilling the Einstein field equations. Under these conditions, each equilibrium configuration is characterized by its baryonic mass, $M_b$, and its angular momentum, $J$. Instead of integrating the Einstein equations at every time step, we adopt the following procedure:
\begin{enumerate}
    \item 
    We neglect any direct effect of the binary companion gravitational field in the self-gravity of the other star. This assumption is based on the relatively large binary separations involved, i.e., about $10^3$--$10^4$ Schwarzschild radii of the NS. Therefore, we calculate the evolution of each NS independently on the other. The only effect caused by the presence of the binary companion on the baryonic mass accretion rate as discussed in Sec. \ref{sec:II} (see Fig. \ref{fig:Mdots_Porb}).
 \item 
 We use the NS equilibrium configurations calculated in \cite{2015PhRvD..92b3007C} through the numerical integration of the Einstein equations with the RNS numerical code. The main result we use here is that the gravitational (Komar) mass of the NS, fully determined by $M_b$ and $J$, i.e., $M = M(M_b, J)$, can be obtained from the following approximately EOS-independent fitting formula relating the above three quantities \cite{2015PhRvD..92b3007C}:
    \begin{equation}\label{eq:MbMns}
  \frac{M_b}{M_\odot} = \frac{M}{M_\odot}+\frac{13}{200}\left( \frac{M}{M_\odot} \right)^2\left(1-\frac{1}{130}j^{1.7} \right),
\end{equation}
where $j\equiv cJ/(GM_\odot^2)$ is the dimensionless angular momentum. We notice that this dimensionless parameter is different from the dimensionless Kerr parameter, $\alpha = c J /(G M^2)$, so the two parameters are related by $j = \alpha (M/M_\odot)^2$.
    \item 
    It has been shown in full generality that the evolution of the NS gravitational mass satisfies \cite{2000AstL...26..772S}
    \begin{equation}\label{eq:Mgrav_Evol}
      \dot{M}=\left( \frac{\partial M}{\partial M_b} \right)_{J} \, \dot{M}_b + \left( \frac{\partial M} {\partial J}\right)_{M_b}\, \dot{J},
\end{equation}
where $\dot{M}_b$ is the baryonic mass accretion rate and $\dot{J}$ is the rate at which angular momentum is transferred to the NS. We recall that $\dot{M}_b$ is here obtained from the SPH numerical simulations described in Sec. \ref{sec:II}. For the numerical integration of Eq. \eqref{eq:Mgrav_Evol}, we must know the partial derivatives which comes out from the integration of the Einstein equations.
    \item 
    We can obtain analytic expressions for the two partial derivatives in Eq. \eqref{eq:Mgrav_Evol} readily from Eq. \eqref{eq:MbMns}:
\begin{align}
    \left(\frac{\partial \mu}{\partial \mu_b} \right)_{j} &= \frac{1}{1+\frac{13}{100}\mu\left(1-\frac{1}{130}j^{1.7}\right)},\\ 
    \left( \frac{\partial \mu} {\partial j}\right)_{\mu_b} &= \frac{\frac{1.7}{2000}\mu^2 j^{0.7}}{1+\frac{13}{100}\mu\left(1-\frac{1}{130}j^{1.7}\right)},
\end{align}
where $\mu = M/M_\odot$.
    \item 
    Finally, we must supply the angular momentum conservation equation by specifying the torques acting onto the stars, i.e., an equation for $\dot{J}$. We assume that the NSs are subjected to the positive torque by accretion and the negative torque due to the magnetic braking mechanism, i.e.
    \begin{equation}\label{eq:Jdot}
        \dot{J} = \tau_{\rm acc} + \tau_{\rm mag},
    \end{equation}
    where each torque is specified below. 
\end{enumerate}
}

{With the above procedure, we obtain the evolution of $M$ and $J$ with time. We now proceed to the specification of the torques acting on the NSs. We start with the angular momentum transfer by accretion.} Accordingly to the numerical simulations, we assume that the infalling material form a disk around the star before being accreted. Therefore, the accreted matter exerts a (positive) torque
\begin{equation}\label{eq:chi}
  \tau_{\rm acc} = \chi\,l\,\dot{M}_b,
\end{equation}
where $l$ the specific (i.e. per unit mass) angular momentum of the innermost stable circular orbit around the NS, and $\chi \leq 1$ is an efficiency parameter  of angular momentum transfer. {The angular momentum of the last stable circular orbit around rotating NSs was calculated in \cite{2017PhRvD..96b4046C} from numerical integration of the geodesic equations in the exterior stationary axially symmetric spacetime of the NSs. The latter were obtained from numerical integration of the Einstein equations for the same EOS used in \cite{2015PhRvD..92b3007C}. The relevant result that we use here is the following approximate EOS-independent expression of $l$ in terms of $M$ and $J$ that fit the numerical results \cite{2017PhRvD..96b4046C}:
\begin{equation}
  l = \frac{G M}{c}\left[2\sqrt{3}\mp 0.37\left( \frac{j}{M/M_\odot} \right)^{0.85}\right],
  \label{eq:lISO}
\end{equation}
where it can be seen that the first term is the specific angular momentum of the last stable circular orbit for the Schwarzschild metric and the second term is a non-linear correction due to the rotation. 
}

The stars are {also} subjected to the (negative) torque by the magnetic field. We adopt the {point\footnote{{We neglect finite-size effects since the ratio between the stellar radius, $R$, and the light-cylinder radius, $R_L = c/\Omega$, is small, i.e., $R/R_L = \Omega R/c \lesssim 0.1$ (see \cite{2015MNRAS.450..714P}, for further details).}}} dipole+quadrupole magnetic field model \cite{2015MNRAS.450..714P}:
{
\begin{align}
    \tau_{\rm mag} &= \tau_{\rm dip} + \tau_{\rm quad},\label{eq:taumag}\\
    \tau_{\rm dip} &= -\frac{2}{3} \frac{B_{\rm dip}^2 R^6 \Omega^3}{c^3} \sin^2{\xi},\\
    \tau_{\rm quad} &= -\frac{32}{135} \frac{B_{\rm quad}^2 R^8 \Omega^5}{c^5} \sin^2\theta_1 (\cos^2\theta_2+10\sin^2\theta_2),
\end{align}
where {$\xi$} is the inclination angle of the magnetic dipole moment with respect to the rotation axis, and the angles $\theta_1$ and $\theta_2$ specify the geometry of the quadrupole field. For the dipole magnetic field, with strength $B_{\rm dip}$, the pure axisymmetric mode $m = 0$ is given by {$\xi = 0$}, and the pure $m=1$ mode by {$\xi = \pi/2$}. For the quadrupole, with strength $B_{\rm quad}$, the $m=0$ mode is given by $\theta_1 = 0$ and any value of $\theta_2$, the $m=1$ mode is given by $\theta_1 = \pi/2$ and $\theta_2=0$, and the $m=2$ mode is set by $\theta_1 = \theta_2 = \pi/2$. For our estimates, we adopt the $m=1$ mode for the dipole and leave the quadrupole free to range between the $m=1$ and $m=2$ modes. Therefore, we can write the total magnetic torque (\ref{eq:taumag}) as
}
\begin{equation}\label{eq:Lsd}
    \tau_{\rm mag} = -\frac{2}{3} \frac{B_{\rm dip}^2 R^6 \Omega^3}{c^3}\left( 1 + \eta^2 \frac{16}{45} \frac{R^2 \Omega^2}{c^2} \right),
\end{equation}
where $\eta$ is the quadrupole to dipole magnetic field strength ratio defined by
\begin{equation}\label{eq:eta}
    \eta \equiv \sqrt{\cos^2\theta_2+10\sin^2\theta_2} \frac{B_{\rm quad}}{B_{\rm dip}}.
\end{equation}

We compute the stellar angular velocity as $\Omega = J/I$, being $I$ the moment of inertia which we estimate with the EOS-independent approximate expression \cite{2019JPhG...46c4001W}
\begin{equation}\label{eq:ILS}
I \approx \left( \frac{G}{c^2} \right)^2 M^3 \sum_{i=1}^4\frac{b_i}{(M/M_\odot)^i},
\end{equation}
where $b_1 = 1.0334$, $b_2 = 30.7271$, $b_3 = -12.8839$, and $b_4 = 2.8841$. {This expression for the moment of inertia neglects the contribution of rotation. We expect an appreciable contribution near the mass-shedding limit where it can be up to $20\%$ larger than a non-rotating estimate (see, e.g., Fig. 3 and related discussion in \cite{2016MNRAS.459..646B}). This could change at some level the quantitative estimates but not the qualitative conclusions.}

Having specified all the above, the evolution of the stellar angular momentum can be computed from angular momentum conservation equation {\eqref{eq:Jdot}, with the aid of Eqs. \eqref{eq:chi} and \eqref{eq:Lsd}.}

\begin{figure*}
    \centering
 \includegraphics[width=\hsize,clip]{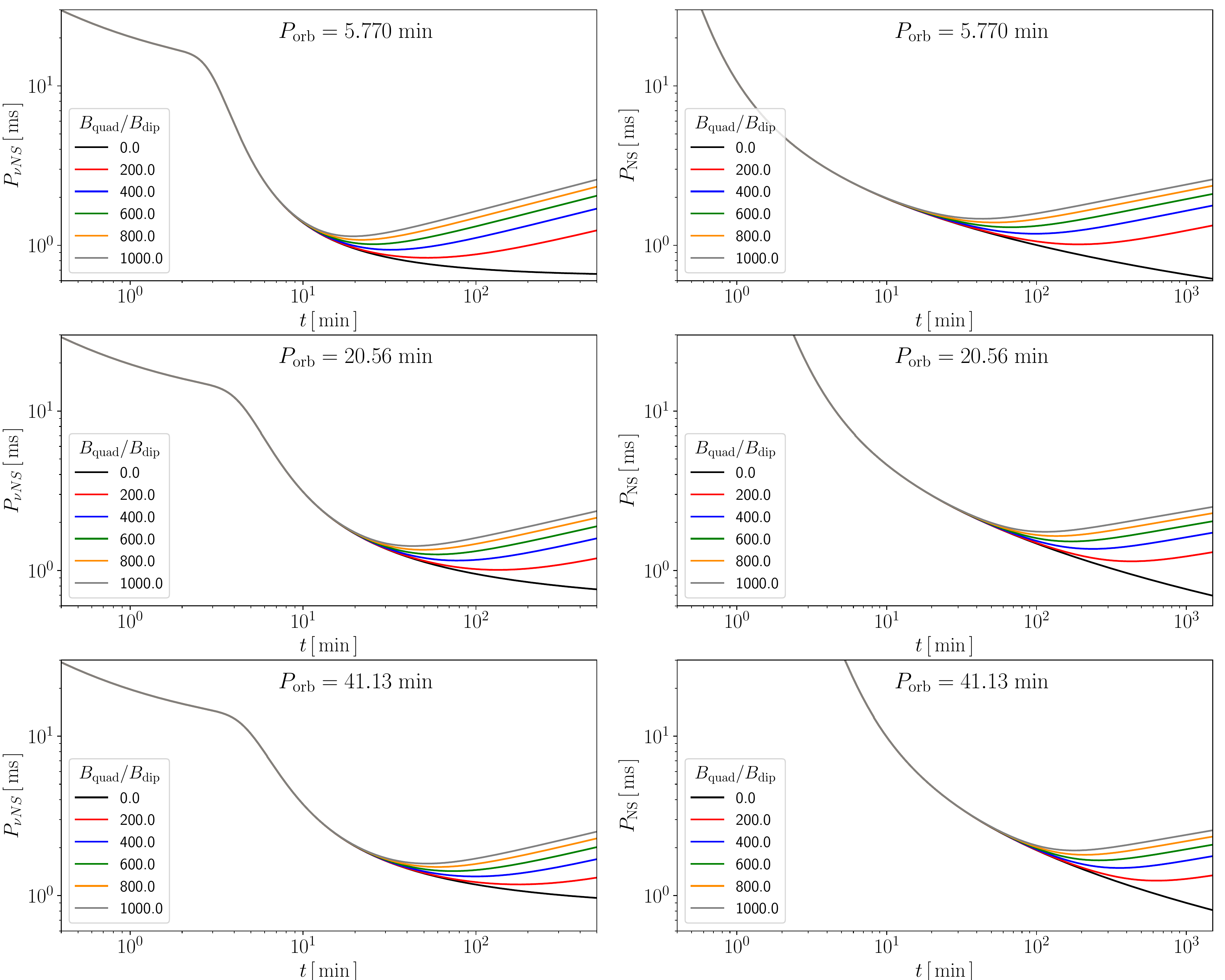}
\caption{{Evolution of the $\nu$NS (left panel) and the NS companion (right panel) rotation period starting from a non-rotating star, i.e., $j(0) = 0$. The $\nu$NS and the NS companion initial mass are $1.8 M_\odot$ and $2 M_\odot$, respectively. For both stars the radius is $10^6$ cm, the magnetic dipole field strength is $B_{\rm dip} = 10^{13}$ G, and the quadrupole to dipole strength ratio is varied from $0$ (pure dipole) to $1000$. The quadrupole field is set in the $m=1$ mode, i.e., $(\theta_1, \theta_2) = (\pi/2, 0)$.} The plot shows the phases of spinup and spindown of the $\nu$NS. The spindown phase is enhanced by the presence of the quadrupole component. The black curves correspond to the case of a pure dipole magnetic field.}\label{fig:PnutBquad}
\end{figure*}

We can now proceed to the integration of the system of differential equations (\ref{eq:Mgrav_Evol}) and (\ref{eq:Jdot}) for the $\nu$NS and the NS companion. For this, we must specify the initial mass and angular momentum, as well as the strength of the magnetic dipole {and quadrupole. We plot in Fig. \ref{fig:PnutBquad} examples} of evolution of the rotation period ($P=2\pi/\Omega$) of the $\nu$NS with time. We have set initially a non-rotating configuration, i.e., $j(0) = 0$, an initial mass of $1.8 M_\odot$, a magnetic dipole strength $B_{\rm dip} = 10^{13}$~G, a stellar radius $R=10^6$ cm, {selected values of the quadrupole to dipole strength ratio, and the $m=1$ mode of the quadrupole, i.e., $(\theta_1, \theta_2) = (\pi/2, 0)$}. The figure shows that for shorter orbital periods the $\nu$NS becomes a faster rotator. The angular momentum transferred by accretion is proportional to the accretion rate, see Eq. (\ref{eq:chi}), so higher accretion rates imply faster rotation rates. Indeed, the presence of the NS companion creates a second peak of accretion onto the $\nu$NS, and that peak of accretion becomes higher for shorter orbital periods (see Fig. \ref{fig:Mdots_Porb}). Therefore, the shorter the orbital period, the more the accretion rate onto the $\nu$NS, and consequently the faster its rotation rate. We can see the phase of spinup of the $\nu$NS followed by the phase of spindown due to magnetic braking. The presence of a non-zero quadrupole component enhances the spindown phase, making it start earlier and at a higher rotation period (the minimum of the curve shifts to upper left values). The black curves correspond to the case of a pure dipole magnetic field.
%

\section{Rotational power evolution}\label{sec:V}

\begin{figure*}
    \centering
 \includegraphics[width=\hsize,clip]{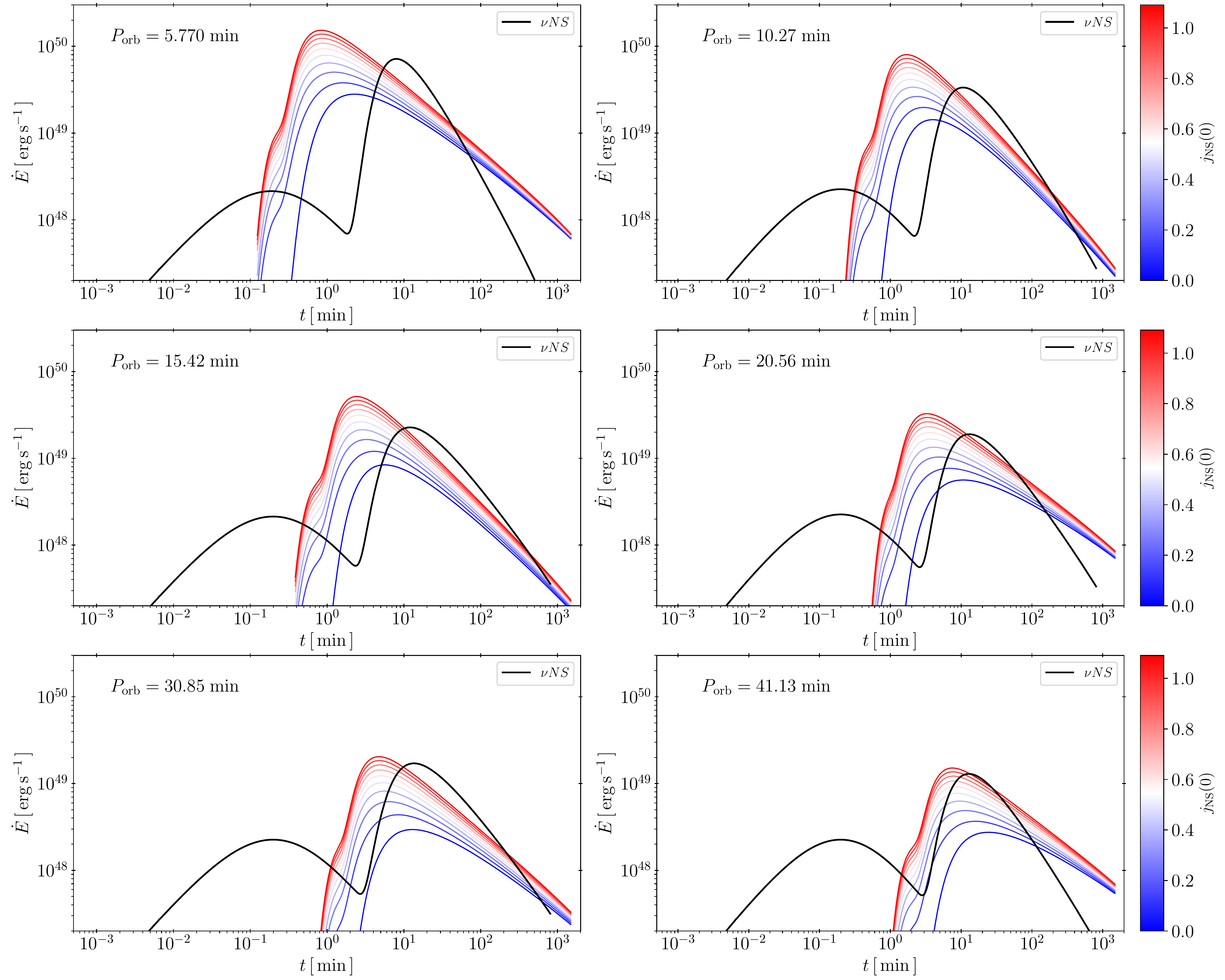}
\caption{Evolution with time of $\dot{E}$, given by Eq. (\ref{eq:Tdot}), for the $\nu$NS and the NS companion. We set the $\nu$NS initially as non-rotating (i.e. $J(0)=0$), and explore the effects of a non-zero initial value of the NS companion angular momentum. In this example, we have initial values of the NS dimensionless angular momentum from $j(0)=0$ (first curve in blue from bottom to top), to $j(0)=1$ (last curve in orange from top to bottom). Each plot corresponds to a binary with a different orbital period (see also Fig. \ref{fig:Mdots_Porb}). In this example, for simplicity, we set the same initial mass and magnetic dipole moment for both stars, respectively, $1.8 M_\odot$, radius $R=10^6$ cm, and a pure dipole ($\eta=0$) with magnetic field strength $B_{\rm dip} = 10^{13}$ G.}
    \label{fig:dTdt}
\end{figure*}

We turn now to estimate the energy gained during the accretion process and that can be released. We plot in Fig. \ref{fig:dTdt} the time derivative of the energy \cite{1969ApJ...157.1395O}
\begin{equation}\label{eq:Tdot}
    \dot{E} = \Omega \dot{J},
\end{equation}
for the $\nu$NS and the NS companion. The energy gain follows the behavior of the accretion rate (see Fig. \ref{fig:Mdots_Porb}), namely, for the $\nu$NS is characterized by two peaks, while the NS companion shows a single peak. The reason for this is that at early times the rotational evolution is dominated by the accretion torque which is proportional to the accretion rate. The relative position and intensity of the peaks depends on the orbital period and on the initial angular momentum of the NSs. In this example, we set an initially non-rotating $\nu$NS, and non-zero values for the initial angular momentum of the NS companion (which is more likely to be fast rotator at the time of the BdHN occurrence) in the range $0\leq j(0)\leq 1$. This corresponds to an initial NS angular momentum in the range $0\leq J(0)\leq 8.8\times 10^{48}$ g cm$^2$ s$^{-1}$. Assuming, for instance, an initial mass of $1.8 M_\odot$, for which Eq. (\ref{eq:ILS}) gives a moment of inertia $I \approx 1.43\times 10^{45}$ g cm$^2$, this leads to a range of initial rotation frequency $0\leq f(0)\leq 685$ {Hz}, i.e., initial rotation periods $P(0)\geq 1.46$ ms. {This is far from the mass-shedding limit. It has been shown that at mass-shedding, the Kerr parameter reaches an EOS-independent maximum value $\alpha \approx 0.7$ (see \cite{2011ApJ...728...12L} and also Fig. 7 and Table I in \cite{2015PhRvD..92b3007C}). Instead, the specific value of the angular momentum (or the rotation frequency) depends on the mass of the configuration that reached the mass-shedding limit. The above NS of $1.8 M_\odot$ NS would be at mass-shedding if $j \approx 0.7\times 1.8^2 \approx 2.27$. For instance, for the TM1 nuclear EOS, the configuration at the crossing point between the mass-shedding limit and the secular axisymmetric instability has $\alpha \approx 0.7$ and $M \approx 2.62 M_\odot$, which leads to $j\approx 4.8$ and a rotation frequency of $1.34$ {kHz} \cite{2015PhRvD..92b3007C}.
}

The second peak of $\dot{E}$ of the $\nu$NS is a unique feature of BdHNe because this is originated by the second peak of fallback accretion onto the $\nu$NS induced by the presence of the NS companion (see Fig. \ref{fig:Mdots_Porb} and \citealp{2019ApJ...871...14B} for further details). The intensity of the second peak of the $\nu$NS decreases for longer orbital periods. In fact, for a single-star system, the second peak disappears as the second peak of fallback accretion vanishes \cite{2019ApJ...871...14B}. The intensity of this peak is comparable to that of the NS companion.

\begin{figure}[ht]
    \centering
 \includegraphics[width=\hsize,clip]{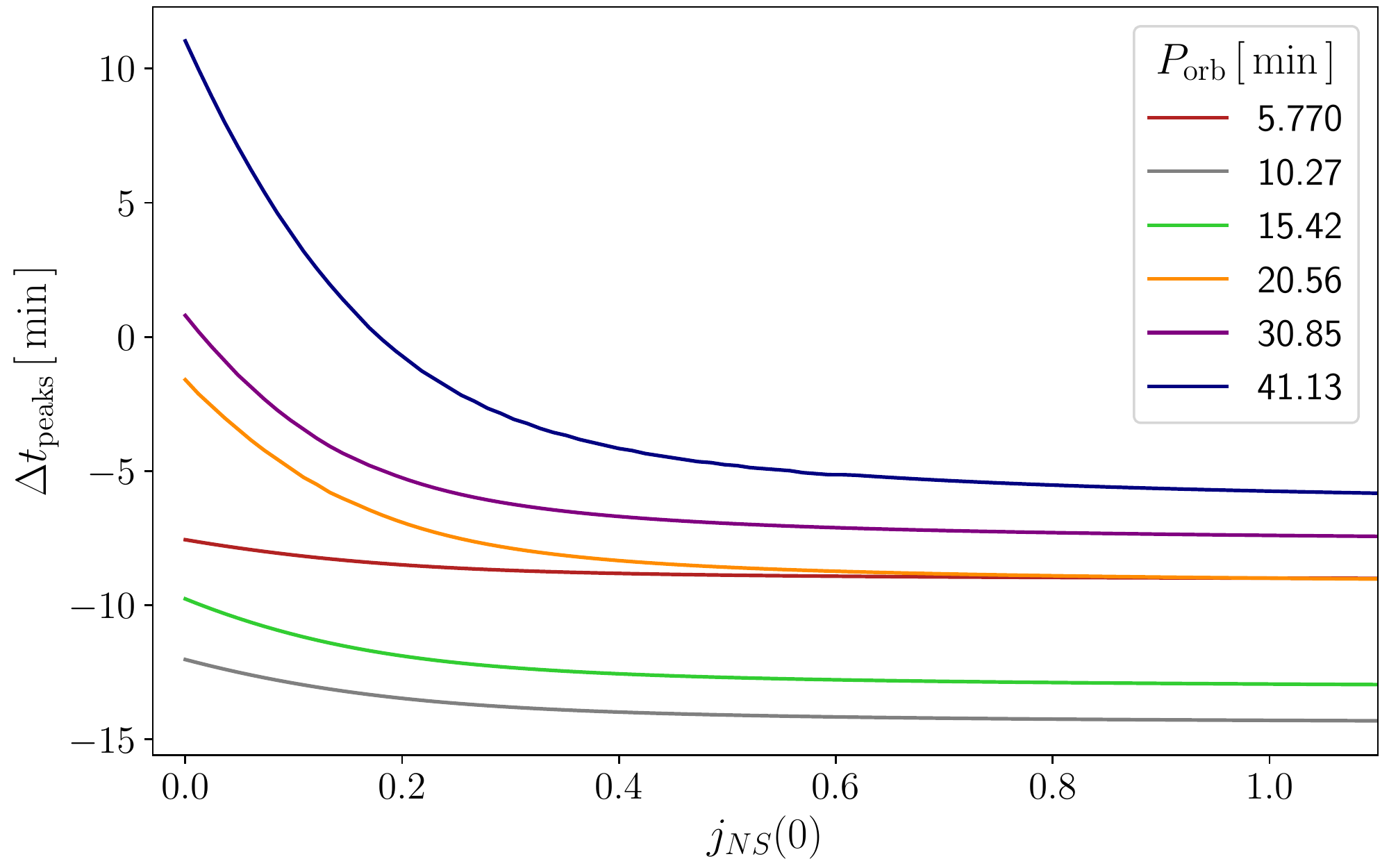}
\caption{Time separation between the peak of $\dot{E}$ of the NS companion and the second peak of $\dot{E}$ of the $\nu$NS (see Fig.\ref{fig:dTdt}).}
    \label{fig:deltaT2}
\end{figure}

For a faster initial rotation rate, the peak of $\dot{E}$ for the NS companion increases its intensity and its time of occurrence, $t_{\rm peak,NS}$, shifts to earlier times. Consequently, the time of occurrence of the $\nu$NS second peak, $t_{\rm peak2,\nu NS}$, and $t_{\rm peak,NS}$ of the NS companion, can be very close to each other (e.g., a few seconds of time separation) or even simultaneous depending on the orbital period and the initial angular momentum of the stars. For the interpretation of GRB data, it is useful to estimate the time separation of these two peaks, i.e.
\begin{equation}
    \Delta t_{\rm peaks} =t_{\rm peak,NS}-t_{\rm peak2,\nu NS}.
\end{equation}
We show in Fig. \ref{fig:deltaT2} $\Delta t_{\rm peaks}$ as a function of the initial (dimensionless) angular momentum of the NS companion, $j(0)$, and for the selected orbital periods relevant for BdHN I and II studied in this work. We have set the $\nu$NS as initially non-rotating. When both stars are initially non-rotating, i.e., for $j(0)=0$, the figure shows that $\Delta t_{\rm peaks}$ starts positive for long orbital periods, so the peak of the NS occurs after the second peak of the $\nu$NS. For shorter orbital periods, the time separation of the peaks decreases. The peaks become simultaneous for an orbital period close to $P_{\rm orb}\approx 31$ min, where $\Delta t_{\rm peaks}$ vanishes. For $P_{\rm orb}\lesssim  31$ min, $\Delta t_{\rm peaks}$ becomes negative, so the peak of the NS occurs before the second peak of the $\nu$NS. If we read Fig. \ref{fig:deltaT2} at a fixed orbital period, we see that the time separation between peaks decreases for larger initial angular momentum of the NS companion. Therefore, only in binaries with $P_{\rm orb}\gtrsim 31$ min the two peaks can occur simultaneously or very close to each other, e.g., with a few seconds of time separation.

\section{Conclusions}\label{sec:VI}

The increasing amount and quality of the multiwavelength data of GRBs allow to test theoretical models in great detail. With the aim of providing further tests of the BdHN scenario of GRBs, we calculated in this article the evolution of a BdHN in the first minutes after the SN explosion. Specifically, we calculated the evolution of the $\nu$NS formed at the center of the SN, and of the NS companion, as a result of the mass and angular momentum transferred onto them by the hypercritical accretion of SN ejecta. We calculated the accretion rate onto the $\nu$NS and the NS companion using three-dimensional SPH numerical simulations performed with the SNSPH code adapted to BdHNe presented in \cite{2019ApJ...871...14B}. 

We followed the evolution of mass and angular momentum calculated from energy and angular momentum conservation accounting for general relativistic effects {by using effective models for the NS binding energy and the specific angular momentum transferred by the accreted matter}. We have compared and contrasted the features of the accretion rate onto the two stars (see, e.g., Fig. \ref{fig:Mdots_Porb}). Particularly relevant is the two-peak structure of the accretion rate onto the $\nu$NS with respect to the single-peak structure of the one onto the NS companion. 

With the knowledge of the accretion rate and having specified the torques due to the accretion process and the magnetic field, we integrated the energy and angular momentum equilibrium equations to determine the rotational evolution of the $\nu$NS and the NS companion. We have shown that the NSs evolve first in a spinup phase dominated by the positive torque due to the accretion process to then start a spindown phase dominated by the negative torque due to the presence of the magnetic field. We considered both a pure magnetic dipole or a dipole+quadrupole magnetic field configurations (see, e.g., Fig. \ref{fig:PnutBquad}). We show that the shorter the orbital period, the higher the rotation rate acquired by the $\nu$NS during the accretion process (see Fig. \ref{fig:PnutBquad}). This result implies that the $\nu$NS in BdHN II are slower rotators than the $\nu$NS in BdHN I.

We then focused on the evolution of the power gained by the NSs during their corresponding accretion processes. This serves as an estimate of the releasable power by the $\nu$NS and the NS companion during this early BdHN evolution. We have shown that the evolution of the power with time reflects the features of the accretion rate, i.e., it has a two-peak structure for the $\nu$NS and a single-peak structure for the NS companion (see Fig. \ref{fig:dTdt}). We have also studied the dependence of the NS power both on the orbital period and the initial angular momentum of the NS companion. The most relevant feature from the observational point of view is that the second peak of the $\nu$NS power is comparable both in intensity as well as in the time of occurrence to the peak of the NS companion power (see Fig. \ref{fig:dTdt}). We deepened into this latter feature by studying the time separation between the second peak of the $\nu$NS power and the peak of the NS companion power, as a function of the initial angular momentum of the NS companion and for a variety of orbital periods (see Fig. \ref{fig:deltaT2}).

The rotational energy powered gained by the $\nu$NS and the NS companion during their early accretion processes, and the accretion power itself, can lead to early emissions prior to the main prompt emission phase. The results of this work imply the possibility of observing precursors in the X-ray and/or in the gamma-rays with a double-peak structure. Therefore, this theoretical prediction of the BdHN evolution and associated emissions during the first minutes after the SN explosion are relevant for the detailed interpretation of GRB multiwavelength data before the prompt emission. {In BdHNe II, low-luminosity GRBs in which the NS companion does not collapse to a BH, the $\nu$NS and NS emissions presented in this article could be observed as a double-peak prompt emission, as shown in the recent analysis of  GRB 190829A \cite{2022ApJ...936..190W}.}

The ultrarelativistic prompt emission (UPE) phase of the GRB {(in a BdHN I)}, the GeV emission, and the following X-optical-radio afterglow are explained by different physical processes occurring in a BdHN after its early evolution analyzed in this work. For details on the UPE phase, see \cite{2021PhRvD.104f3043M}, for the GeV emission, see \cite{2019ApJ...886...82R, 2020EPJC...80..300R, 2021A&A...649A..75M, 2022ApJ...929...56R}, and for the afterglow, see \cite{2018ApJ...869..101R, 2019ApJ...874...39W, 2020ApJ...893..148R}. Therefore, the theoretical work presented in this article complements the self-consistent picture developed in the BdHN model for a complete interpretation of the multiwavelength data of long GRBs, starting from the X and gamma-ray precursors, to the UPE in the MeV, to the GeV emission, to the X-optical-radio afterglow, and finally to the optical emission powered by the radioactive decay of nickel in the SN ejecta.

\begin{acknowledgments}
{
We thank the anonymous referees for the insightful comments and suggestions that helped us to improve the presentation of the article. 
} L.M.B. is supported by the Vicerrector\'ia de Investigación y Extensi\'on - Universidad Industrial de Santander Postdoctoral Fellowship Program No. 2022000293. 
\end{acknowledgments}


%

\end{document}